\documentclass[11pt]{article}

\usepackage[margin=1in]{geometry}
\usepackage{amsmath,amssymb}
\usepackage{graphicx}
\usepackage{booktabs}
\usepackage{xcolor}
\usepackage[backend=biber, style=numeric, sorting=none]{biblatex}
\DeclareCiteCommand{\citenum}
  {\usebibmacro{prenote}}
  {\printtext[bibhyperref]{\printfield{labelnumber}}}
  {\multicitedelim}
  {\usebibmacro{postnote}}
\usepackage[colorlinks=true,linkcolor=blue,citecolor=blue,urlcolor=blue]{hyperref}

\newcommand{\code}[3]{[\![#1,#2,#3]\!]}

\title{First fault-tolerant quantum memory demonstration for a generalized superfast encoding}

\author{James Brown and Kenny Heitritter\\ \normalsize qBraid Co., Chicago, IL, United States}

\date{}

\begin{document}
\maketitle

\begin{abstract}
The Generalized Superfast Encoding (GSE) is a fermion-to-qubit mapping that has error-correcting/detecting properties. To this point, all demonstrations have been relegated to error-detection only, as no fault-tolerance under circuit-level noise has been observed.  Here, we introduce an even-distance $d$ constant stabilizer-weight GSE where each of $N$ modes is assigned a $d$-qubit block arranged on a ring. The resulting stabilizer generators have constant weight 4 or 6 for any even distance $d$. Furthermore, the full stabilizer set of this construction can always be partitioned into four qubit-wise commuting groups, which enables compact syndrome-extraction scheduling. We simulate quantum memory experiments under circuit-level depolarizing noise for two instances of this code, $\code{48}{8}{6}$ and $\code{64}{8}{8}$ and the threshold is observed to be  $\approx 4\times10^{-3}$. This is, to our knowledge, the first fault-tolerant quantum-memory characterization of a fermion-mapping with threshold-like scaling.
\end{abstract}

\section{Introduction}

Simulating fermionic systems on a quantum computer requires an encoding of fermionic modes into qubits. Canonical choices (such as the Jordan--Wigner transformation~\cite{jordanwigner1928} and the Bravyi--Kitaev encoding~\cite{bravyikitaev2002}) map $m$ fermionic modes onto $m$ qubits which generally produce nonlocal and high-weight Pauli operators\cite{bravyikitaev2002, Seeley_2012}. Graph-based Fermion encodings instead place qubits on the edges of the Fermionic interaction graph\cite{Derby_2021, Verstraete_2005, Ball_2005, Miller_2023, JKMN, Vlasov_2022, setia2019superfast, landahl2023logicalfermionsfaulttolerantquantum, chiew2025optimalfermionqubitmappingsquadratic}. This trades a (potentially) modest increase in qubit count for operators whose weight can be much lower for the Hamiltonians these graphs are derived from. The Generalized Superfast Encoding (GSE) of Setia, Bravyi, Mezzacapo, and Whitfield~\cite{setia2019superfast} can also generate stabilizers that can be used for error mitigation and correction. For example, a suitable choice of local Majorana operators can correct all single-qubit errors for a graph with interaction degree greater than 6\cite{setia2019superfast}, at no extra qubit cost over the original superfast encoding\cite{Setia_2018}.

Building on this, Ref~\citenum{brown2025efficient} developed a suite of techniques that make GSE competitive with the Jordan-Wigner and Bravyi-Kitaev mappings for realistic molecular Hamiltonians. Additionally, a multi-edge construction for a ring interaction graph was introduced that increases code distance with fixed stabilizer weight. The stabilizer weight was $6$ for any odd code distance $d = 2k+1$. Up to this point, no GSE construction (or any of the other graph-like encodings) has yet been characterized as a genuine quantum memory. That is, by showing that the logical error rate decreases exponentially with decreasing physical error rate under realistic circuit-level noise.


In this work, we show that an even-distance version of the ring-geometry GSE of Ref \citenum{brown2025efficient} has constant stabilizer weight of 4 or 6. We also show that this stabilizer set always partitions into exactly 4 mutually qubit-wise-commuting groups. This makes the depth of any syndrome-extraction schedule also independent of code distance. To generate the syndrome extraction circuit, we use AlphaSyndrome~\cite{liu2026alphasyndrome}. AlphaSyndrome is used to generate decoder-aware bare-ancilla syndrome-extraction circuits for two instances of this code family. The logical memory errors are decoded under circuit-level-noise simulations with the Relay-BP~\cite{muller2025relaybp} decoder and show threshold like scaling for memory experiments.

Our contributions are:
\begin{enumerate}
  \item An even-distance GSE construction with stabilizers of constant weight 4 or 6 for any even $d$, whose stabilizer set is always 4-colorable in qubit-wise commuting groups (Section~\ref{sec:construction}).
  \item The first quantum-memory simulation of any fermion-to-qubit encoding under circuit-level noise that achieves threshold-like scaling near $p\approx4\times10^{-3}$ (Section~\ref{sec:results}).
  \item A transversal code-switch procedure for which all single-particle logical operations have weight at most $2d+(N-2)/2$.
\end{enumerate}

The remainder of the paper is organized as follows. Section~\ref{sec:background} reviews GSE, the prior odd-distance construction, syndrome-extraction scheduling, and Relay-BP. Section~\ref{sec:construction} presents the even-weight construction. Section~\ref{sec:methods} describes the circuit synthesis and decoding methodology. Section~\ref{sec:results} presents the memory simulation results. Section~\ref{sec:discussion} discusses limitations and Section~\ref{sec:future} outlines future work.

\section{Background}
\label{sec:background}

\subsection{The Generalized Superfast Encoding}

The GSE maps a system of $m$ Fermionic modes interacting on a graph $G=(V,E)$ to $n$ qubits by placing $d(i)/2$ qubits at each vertex $i \in V$, where $d(i)$ is the vertex degree. At each vertex, a set of local Majorana operators $\gamma_{i,1},\ldots,\gamma_{i,d(i)}$ is chosen to generate the full Pauli group on that vertex's qubits, subject to the usual Majorana anticommutation relations. Edge operators $\tilde{A}_{j,k} = \epsilon_{j,k}\gamma_{j,p}\gamma_{k,q}$ and vertex (occupation) operators $\tilde{B}_i = (-i)^{d(i)/2}\gamma_{i,1}\cdots\gamma_{i,d(i)}$ reproduce the algebra of Fermionic edge and number operators once restricted to a stabilized code space. That code space is defined by minimum cycle basis of loop operators $\tilde{A}(\zeta)$ of the interaction graph $G$. We use the standard notation $\code{n}{k}{d}$ for a code encoding $k$ logical qubits into $n$ physical qubits at distance $d$.

\subsection{Prior constant-weight GSE construction: the odd-distance multi-edge ring}
\label{sec:oddring}

Brown et al.~\cite{brown2025efficient} showed that starting from a simple ring graph connecting $N$ modes and adding parallel (multi-)edges between neighbors increases code distance without increasing stabilizer weight. Choosing local Majorana orderings
\[
m = (\underbrace{Z\cdots Z}_{k}\, Y\, \underbrace{I\cdots I}_{k}), \qquad
n = (\underbrace{Z\cdots Z}_{k}\, X\, \underbrace{I\cdots I}_{k}),
\]
and cyclic-permutation edge operators $A^c_{i,j}$ built from $c$-shifted copies of $m,n$, the stabilizers $S_{i,c} = A^c_{i,i+1}A^{c+1}_{i,i+1}$ formed from the multi-edges have weight exactly $6$. This construction is restricted to odd $d$ by the structure of the $m,n$ orderings, which place a single $Y$ (or $X$) flanked symmetrically by $k$ $Z$'s and $k$ identities. This is not the first time that a local fermion-to-qubit mapping has constant stabilizer weight with increasing distance\cite{Algaba2025, Wei2025}.

\subsection{Bare-ancilla syndrome extraction and circuit scheduling}

A standard, hardware-agnostic, realization of stabilizer measurement is \emph{bare-ancilla} syndrome extraction. For each stabilizer generator $S = \sigma_n\cdots\sigma_1$, prepare a dedicated ancilla in $|+\rangle$, apply a sequence of entangling gates with appropriate single-qubit basis-change gates on the data qubits and measure the ancilla in the $X$ basis. Because stabilizers commute, many orderings and parallelizations of these per-stabilizer check sequences are logically equivalent in the absence of noise. However, only qubit-wise commuting groups can be measured simultaneously without completing the previous stabilizer measurement. In CSS codes (where all stabilizers are either all $X$ or all $Z$ operators), at most two layers of stabilizer measurements are required. For non-CSS codes (where stabilizers are general Pauli operators), a linear number of layers can be required. Also, under realistic noise, a single fault occurring partway through a multi-qubit check circuit (a \emph{hook error}\cite{Dennis_2002}) can propagate to a correlated, higher-weight error on the data register. We used AlphaSyndrome~\cite{liu2026alphasyndrome} to generate the schedule and reduce the impact of hook errors. Further work will examine whether hand-derived schedules can provide improved circuit depth (using techniques of Ref \citenum{Tremblay_2022}) or provide a better starting point for Alphasyndrome.

\subsection{Belief propagation and Relay-BP}

Belief propagation (BP) decodes a circuit-level stabilizer measurement by iteratively passing log-likelihood messages between error and check nodes of the decoding graph induced by the circuit. Vanilla BP is fast and naturally parallel, but frequently fails to converge because of oscillating beliefs\cite{poulin2008iterativedecodingsparsequantum}. Relay-BP~\cite{muller2025relaybp} addresses this by (i) introducing a per-node memory term that mixes each iteration's belief with the running marginal, with memory strength $\gamma_j$ drawn from a disordered (and, crucially, sometimes \emph{negative}) distribution to break these symmetries, and (ii) chaining multiple such runs (``legs'') into a relay, each leg initialized from the previous leg's final marginals, returning the lowest-weight valid correction found across all legs. This combination achieves accuracy comparable to or exceeding BP+OSD and minimum-weight perfect matching on bivariate-bicycle and surface codes while remaining a lightweight, FPGA-amenable message-passing algorithm. 

\section{An even-weight Generalized Superfast Encoding}
\label{sec:construction}

We construct a GSE code for $N$ logical modes, each carrying $d$ physical qubits arranged on a ring, for any \emph{even} distance $d$. We label the $d$ qubits at vertex $i$ as $i\cdot d, i\cdot d+1,\ldots, i\cdot d+(d-1)$, so that the total qubit count is $n = N\cdot d$.

\subsection{Stabilizer generators}

At each vertex we define three families of local Pauli-string patterns of length $d$, built from an $XY$ (or $YX$) pair together with an isolated $Z$ and identity padding:
\begin{align}
\text{``a'' family:} &\quad XY\,\underbrace{I\cdots I}_{(d-2)/2}\,Z\,\underbrace{I\cdots I}_{(d-4)/2}\ ,\qquad YX\,\underbrace{I\cdots I}_{(d-2)/2}\,Z\,\underbrace{I\cdots I}_{(d-4)/2}\ , \\
\text{``middle'':} &\quad \underbrace{I\cdots I}_{(d-2)/2}\,XY\,\underbrace{I\cdots I}_{(d-2)/2}\ ,\qquad \underbrace{I\cdots I}_{(d-2)/2}\,YX\,\underbrace{I\cdots I}_{(d-2)/2}\ , \\
\text{``b'' family:} &\quad Z\,\underbrace{I\cdots I}_{(d-2)/2}\,XY\,\underbrace{I\cdots I}_{(d-4)/2}\ ,\qquad Z\,\underbrace{I\cdots I}_{(d-2)/2}\,YX\,\underbrace{I\cdots I}_{(d-4)/2}\ .
\end{align}
Cyclically shifting the ``a'' pattern by $i = 0,\ldots,d/2-2$ positions within the vertex and concatenating the $XY$- and $YX$-based copies gives $d/2-1$ weight-6 stabilizer generators; the ``middle'' pattern (unshifted) gives a single weight-4 generator; and cyclically shifting the ``b'' pattern gives a further $d/2-1$ weight-6 generators. Each vertex therefore contributes
\[
\left(\frac{d}{2}-1\right) + 1 + \left(\frac{d}{2}-1\right) \;=\; d-1
\]
stabilizer generators, all of weight 4 or 6, and this pattern is replicated by cyclic roll across all $N$ vertices to give the full stabilizer group. The logical vertex $B_i$ operators are the weight-$d$ strings $Z^{\otimes d}\otimes I^{\otimes (N-1)d}$ rolled around the ring. The above does not fully define the code-space with one instance of a path around the loop required. Any $N$ repetitions of the cyclic shifts of
\begin{equation}
    \quad Z\,\underbrace{I\cdots I}_{(d-1)}
\end{equation}
can be used. This is why this code family requires $N>d$ as otherwise, the above logical operation would have weight less than $d$. The above stabilizer fixes the parity of the mapping. If a minus of the above was used, the codes space would have an odd number of electrons. 

We verified this construction directly for $N=8,d=6$ (a $\code{48}{8}{6}$ code) and $N=8,d=8$ (a $\code{64}{8}{8}$ code) by explicit stabilizer tableau construction. We report distance empirically for these two instances; a general closed-form distance formula as a function of $d$ is left to future work.

\subsubsection{Other logical operations}
The connection between the Fermionic operators (e.g. $a_{i}^{\dagger}a_j$) and the vertex and edge operators is fully derived in Ref \citenum{Setia_2018}. Some examples are:
\begin{equation}\label{eq.num}
    a_{i}^{\dagger}a_i = \frac{1-B_i}{2},
\end{equation}
\begin{equation}\label{eq.exc}
    a_{i}^{\dagger}a_j+a_{j}^{\dagger}a_i= \frac{i}{2}\left(A_{ij}B_j+B_iA_{ij}\right),
\end{equation}
and
\begin{equation}\label{eq.ucc}
    a_{i}^{\dagger}a_j-a_{j}^{\dagger}a_i=\frac{i}{2}\left(A_{ij}+B_i A_{ij} B_j\right).
\end{equation}
For this mapping, Eq\ref{eq.num} and Eq\ref{eq.ucc} have weight at most $d+\frac{N-2}{2}$. 

The $Z^{\otimes d}\otimes I^{\otimes (N-1)d}$ operators define the logical $B_i$ operations. The other required operations are $A_{ij}$ operations. Any Fermion operators are products of $B_i$ and $A_{ij}$ operators. For the definition of the local Majoranas above, we obtain the $A_{ij}$ operators as
\begin{align}
\text{``a'' family:} &\quad \,\underbrace{Z\cdots Z}_{d/2}\,Y\,\underbrace{I\cdots I}_{(d-2)/2}, \quad \underbrace{Z\cdots Z}_{d/2}\,X\,\underbrace{I\cdots I}_{(d-2)/2} \\
\text{``b'' family:} &\quad \,\underbrace{Z\cdots Z}_{d/2}\,Y\,\underbrace{I\cdots I}_{(d-2)/2}, \quad \underbrace{Z\cdots Z}_{(d-2)/2}\,X\,\underbrace{I\cdots I}_{d/2} .
\end{align}
Once again, a cyclic permutation is performed within each mode-block and concatenated together. One can see that the ``b'' family has weight-($d+2$) and the ``a'' family has weight-d. However, if you multiply the ``a'' family with the corresponding $B$ operator, for each $i,i+1$, the logical operators become $\quad \,\underbrace{I\cdots I}_{d/2}\,X\,\underbrace{Z\cdots Z}_{(d-2)/2}, \quad \underbrace{Z\cdots Z}_{d/2}\,Y\,\underbrace{Z\cdots Z}_{(d-2)/2}$ which also have weight-d. Therefore, for nearest neighbor $i,i+1$ hoping operations, the operations $A_{i,i+1}, B_iA_{i,i+1}B_{i+1}$ are all weight-d. You can also see that these two operations act on disjoint sets of qubits which can increase parallelism\cite{Bringewatt2023parallelization}. For example, $A_{i,i+1}$ acts on the first $d/2$ qubits of each $i$ block while $B_iA_{i,i+1}B_{i+1}$ acts on the second $d/2$ qubits of each block.

When performing operations between unconnected modes (i.e. $A_{i,j\neq i\pm1}$), one can multiply $A_{i,k}A_{k,l}...A_{z,j}$. When this is done, the weight on each intermediate set of $B_{k}$ qubits is only 1. As the graph is a loop, the furthest distance that two modes can be apart is $\frac{N-2}{2}$. Hence, the largest possible weight for these terms is $d+\frac{N-2}{2}$ 

\begin{equation}
A_{0,1}\left[\prod_{i=1}^{N-2}A_{i,i+1}\right]A_{N-1,N}=\underbrace{Z\cdots Z}_{d/2}\,Y\,\underbrace{I\cdots I}_{(d-2)/2}\prod^N \left[\underbrace{I\cdots I}_{(d-2)/2}\,Z\,\underbrace{I\cdots I}_{d/2} \right] \, \underbrace{Z\cdots Z}_{(d-2)/2}\,X\,\underbrace{I\cdots I}_{d/2},
\end{equation}
which has weight $d+\frac{N-2}{2}$

\subsubsection{Transversal swapping between mappings.}
We now show that the Eq. \ref{eq.exc} operators can also be represented with weight $d+\frac{N-2}{2}$. The choice of local majoranas above can be modified without changing the distance of the code but it does change the weight of the logical operators. One can also equivalently utilize 
\begin{align}
\text{``a'' family:} &\quad \underbrace{I\cdots I}_{(d-4)/2}\,Z\,\,\underbrace{I\cdots I}_{(d-2)/2}XY\ ,\qquad YX\,\underbrace{I\cdots I}_{(d-2)/2}\,Z\,\underbrace{I\cdots I}_{(d-4)/2}\ , \\
\text{``middle'':} &\quad \underbrace{I\cdots I}_{(d-2)/2}\,XY\,\underbrace{I\cdots I}_{(d-2)/2}\ ,\qquad \underbrace{I\cdots I}_{(d-2)/2}\,YX\,\underbrace{I\cdots I}_{(d-2)/2}\ , \\
\text{``b'' family:} &\quad \underbrace{I\cdots I}_{(d-4)/2}\,XY\,\,\underbrace{I\cdots I}_{(d-2)/2}Z\ ,\qquad Z\,\underbrace{I\cdots I}_{(d-2)/2}\,YX\,\underbrace{I\cdots I}_{(d-4)/2}\ .
\end{align}
The transformation between the above description and the initial mapping can be performed by flipping the ordering of the qubits in every even register and then applying an $S$ gate to every qubit in that even register. For example, for a \code{16}{4}{4} would utilize the circuit SWAP $(0,3), (1,2), (8,11), (9,10)$, S $0,1,2,3,8,9,10,11$.

The difference that occurs is that now possible to select $A_{ij}B_i$ and $A_{ij}B_j$ operators that have weight $d+\frac{N-2}{2}$ instead of $B_i A_{ij} B_j$ and $A_{ij}$. Therefore, one can switch between the two codes and always have operators of weight $d+\frac{N-2}{2}$.

\subsection{Four qubit-wise commuting groups}

A key structural property of the stabilizers for this code is that it is always possible to form four qubit-wise commutating groups. Each group can, in principle, be measured within a single round. Since every generator has weight at most 6, the depth of the circuit needed for a syndrome-extraction round is naively $4\times 6 = 24$. The schedule actually synthesized by AlphaSyndrome discussed in Section~\ref{sec:scheduling} can be deeper. $a$ and $b$ families can be partitioned into two qubit-wise commuting groups each with alternating labels depending on the vertices. For even vertices, the order is $1,2,1,2,...$ for the $a$ family and $3,4,3,4,...$ for the $b$-family. For odd vertices it is $2,1,2,1,...$ and $4,3,4,3,...$. The "middle" family can be placed in group 2 or 4 for even vertices and 1 or 3 for odd vertices. 

\section{Methods}
\label{sec:methods}

\subsection{AlphaSyndrome-scheduled syndrome extraction}
\label{sec:scheduling}

For each code instance we synthesized, a bare-ancilla syndrome-extraction is scheduled with AlphaSyndrome\cite{liu2026alphasyndrome}. We use a two-parameter depolarizing noise model with idling error rate $p/10$ and single- and two-qubit gate error rate $p$, where $p$ is the swept physical error rate reported in Section~\ref{sec:results}.

The synthesized schedules are consistent with, but do not exactly saturate, the structural 4-color/weight-6 bound of $24$ ticks derived in Section~\ref{sec:construction} as AlphaSyndrome optimizes for decoder-conditioned logical error rate rather than minimal depth.

\subsection{Relay-BP decoding}
\label{sec:decoding}

Each simulated circuit's detector error model was constructed via Stim and decoded with Relay-BP. The decoder parameters used to produce the reported data are $\gamma_0 = 0.005$, memory-strength interval $\gamma \in [-0.14, 0.16]$, $80$ relay legs (\texttt{num\_sets}), $80$ pre-iterations, and $\texttt{stop\_nconv}=20$, with a per-leg iteration cap of $1000$ for the $\code{48}{8}{6}$ code and $2000$ for the $\code{64}{8}{8}$ code. Sampling was performed via \texttt{sinter}, with $d$ noisy syndrome-extraction rounds per shot.

\section{Results}
\label{sec:results}

\subsection{Experimental setup}

We swept the physical depolarizing error rate $p$ over ten log-spaced points from $2\times10^{-4}$ to $2\times10^{-3}$ for the $\code{48}{8}{6}$ code, and over the same range excluding the lowest point for the $\code{64}{8}{8}$ code, which we omit from the reported curve owing to insufficient statistics (zero or near-zero observed logical failures at that point given the sampled shot budget). For each $(p, d)$ pair we report the logical error rate per round per logical qubit,
\[
\bar{\epsilon}_L(p) = \frac{\#\text{logical failures}}{\#\text{shots}\times k \times d},
\]
with $k=8$ logical modes for both codes and $d$ equal to the code distance (6 or 8).

\subsection{Memory experiment}

\begin{figure}[t]
\centering
\includegraphics[width=0.7\textwidth]{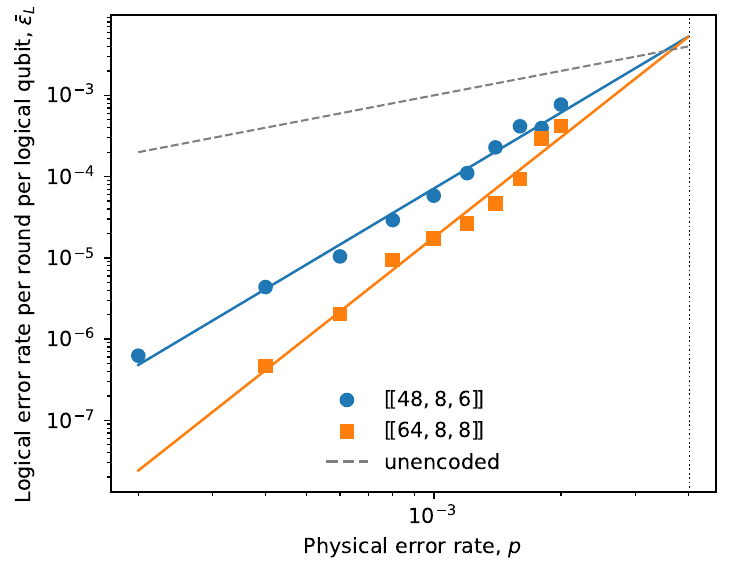}
\caption{Logical error rate per round per logical qubit versus physical depolarizing error rate $p$, for the $\code{48}{8}{6}$ and $\code{64}{8}{8}$ even-weight GSE codes, decoded with Relay-BP using AlphaSyndrome-scheduled syndrome extraction. Solid curves are log-log power-law fits to each code's data. The dashed line is the unencoded (physical) error rate for reference. The higher-distance code suppresses logical error more strongly at low $p$ with the two curves approaching each other near $p\approx4\times10^{-3}$.}
\label{fig:key_plot}
\end{figure}

Figure~\ref{fig:key_plot} shows $\bar{\epsilon}_L(p)$ for both codes on log-log axes, together with power-law fits (linear fits in log-log space) and the unencoded (physical) error rate as a reference line. Over the full sampled range, the higher-distance $\code{64}{8}{8}$ code exhibits a lower logical error rate than the $\code{48}{8}{6}$ code. Extrapolating the two fitted power laws, the curves cross near $p\approx4\times10^{-3}$. This is consistent with a pseudo-threshold on the order of $p\approx4\times10^{-3}$. 

We emphasize two caveats in interpreting this as threshold-like behavior. First, only two code distances were tested. A precise threshold estimate would require at least a third distance. Second, the lowest-$p$ point for the $\code{64}{8}{8}$ code was excluded due to insufficient statistics and very long runtimes (of the order of days). We report the crossing as an order-of-magnitude estimate rather than a precise value for this reason. Within these caveats, the fact that a threshold-like crossing is observed suggests that GSE can be a practical target for fault-tolerant quantum computing of molecules. This threshold value is similar to the recently developed non-CSS mirror codes\cite{khesin2026mirrorcodeshighthresholdquantum}. Therefore, our construction also supports the notion that non-CSS codes can have attractive properties and should be further explored.

\section{Discussion}
\label{sec:discussion}

We introduced an even-distance analogue of the constant-weight multi-edge GSE construction of~\cite{brown2025efficient} with weight 4 or 6 stabilizers for any even $d$. Together with the pre-existing odd-distance construction, this gives constant, low-weight GSE stabilizers across both parities of code distance. Combined with AlphaSyndrome-scheduled syndrome extraction and Relay-BP decoding, we obtained (to our knowledge) the first characterization of a local fermion encoding as a genuine quantum memory. The reported pseudo-threshold should be read as a qualitative, order-of-magnitude estimate rather than a rigorously fit threshold value.

\section{Future work}
\label{sec:future}

A first next step could be a hardware demonstration of this construction. This could include extending the initial error-detecting exploration of Ref~\citenum{brown2026clinr} to a code distance large enough to probe genuine threshold behavior. We will also simulate additional code distances to obtain a proper finite-size-scaling threshold estimate with uncertainty quantification. This will require improved decoding as the runtimes were already lengthy here. To improve the threshold behavior, we will also examine syndrome extraction circuits beyond bare-ancilla\cite{khesin2026mirrorcodeshighthresholdquantum}.

A very important step is extending the construction to support logical operations (not just memory). This can possibly be done using and extending the techniques of Ref \citenum{brown2026clinr} using error correction instead of detection. This first attempt at this would be using the techniques combined with CliNR\cite{DelfosseCliNR} as initially explored in Ref \citenum{brown2026clinr}. This could include adding some ideas from Knill error-correction\cite{murphy2026simplifiedcircuitleveldecodingusing}. 
This exploration could be automated by extending the GSE exploration with AlphaEvolve\cite{novikov2025alphaevolvecodingagentscientific} work of Ref \citenum{heitritter2026evolvingquantumerrorcorrectingencodings}.

\printbibliography

\end{document}